\documentstyle[aps,draft]{revtex}

\def\be{\begin{equation}}
\def\bea{\begin{eqnarray}}
\def\bma{\begin{mathletters}}
\def\ee{\end{equation}}
\def\eea{\end{eqnarray}}
\def\ema{\end{mathletters}}

\begin{document}
\author{Vlatko Vedral}
\title{Landauer's erasure, error correction and entanglement }
\address{Centre for Quantum Computation, Clarendon Laboratory, University of Oxford,\\
Parks Road OX1 3PU}
\date{\today}
\maketitle

\begin{abstract}{Landauer's erasure, thermodynamics, classical and 
quantum error correction, entanglement}
Classical and quantum error correction are presented in the form of
Maxwell's demon and their efficiency analyzed from the thermodynamic point
of view. We explain how Landauer's principle of information erasure applies
to both cases. By then extending this principle to entanglement
manipulations we rederive upper bounds on purification procedures thereby
linking the ''no local increase of entanglement'' principle to the Second
Law of thermodynamics.
\end{abstract}

\section{\protect\bigskip Introduction}

Landauer (1961) showed that any erasure of information is accompanied
by an appropriate increase in entropy. This result was then used by Bennett 
(1982) to finally exorcise Maxwell's demon in a Szilard-like set-up 
(Szilard 1929). His main conclusion was that the increase in entropy is not
necessarily a consequence of observations made by the demon, but accompanies
the resetting of the final state of the demon to be able to start a new
cycle. In other words, information gained has to eventually be erased, which
leads to an increase of entropy in the environment and prevents the Second
Law of thermodynamics from being violated. In fact, the entropy increase in
erasure has to be at least as large as the initial information gain.
Bennett's analysis was, however, completely classical. Soon after this,
Zurek (1984) analyzed the demon quantum mechanically confirming
Bennett's results and since then there has been a number of other related
works on this subject (e.g. Lubkin 1987 and Lloyd 1997). Here, however, we
want to relate the notion of information erasure to the concept of quantum
entanglement. Quantum theory tells us that a measurement process is, in
fact, the creation of correlations (entanglement) between the system under
observation and a measuring apparatus. Loosely speaking, the amount of
entanglement tells us how much information is gained by the apparatus and
therefore it seems natural to assume that Landauer's principle has
implication on entanglement manipulations. In this paper we show that this
indeed is the case; we will argue that Landauer's erasure entropy limits the
increase in the amount of entanglement (more precisely, as will be seen later,
it limits the average increase) between two quantum subsystems when
each one is manipulated separately. This will then lead to an entirely new derivation of
known entanglement measures such as the relative entropy of entanglement 
(Vedral $\&$ Plenio 1998) and the entanglement of creation (Bennett 1996$b$) 
(for a review see Plenio $\&$ Vedral 1998). However, in order to become more 
familiar with Landauer's
principle we first analyze classical error correction and compare it to
quantum error correction using the concept of erasure of information.

\bigskip

The paper is organized as follows. In the next section we demonstrate formal
analogies between classical error correction and a thermodynamical cycle in
general. We show exactly how Landauer's principle is manifested in such a
protocol. In section 3 we repeat this analysis using quantum error
correction and derive the most general statement for information erasure. In
section 4 we apply Landauer's principle to explain recent measures of
entanglement, and link it to the principle of ''no increase of entanglement
by local means''. This is then discussed from two different points of view,
the individual and the ensemble, and a number of different questions is raised.
Finally we discuss the implications of this work to other phenomena in
quantum information theory and state other open problems implicated by our
investigations.

\bigskip

\section{Classical Error Correction and Maxwell's demon}

\bigskip \noindent

We use a simple reversible cycle which is a slight modification of Bennett's
(1982) version of Maxwell's demon to illustrate the process of error
correction. In order to link information to thermodynamics we will use a box
containing a single atom as a representation of a classical bit of
information: the atom in the left hand half (LHH) will represent a $0$, and
the atom in the right hand half (RHH)\ will be a $1$. Now, if the atom is
already confined to one of these halves, and we expand it isothermally and
reversibly to occupy the whole volume, then the entropy of the atom increases
by $\Delta S=k\log 2$ and the free energy decreases by $\Delta F=-kT\log 2.$
The atom does work $\Delta W=kT\log 2.$ Suppose that initially we want to
have our atom in one of the halves in order to be able to do some work as
described. However, suppose also that there is a possibility of error,
namely the atom has a chance of $1/2$ to jump to the other half. Once this
happens we cannot extract any work until we return the atom \ to the initial
state. But this itself requires an amount of work of $kT\log 2$ in an
isothermal compression. We would thus like to be able to correct this error
and so we introduce another atom in a box to monitor the first one. This is
represented in Fig. 1. and the whole error correction protocol goes through
5 stages. \ 



\begin{enumerate}
\item  Initially the atoms are in LHH and RHH of respective boxes.

\item  Then an error happens to atom $A$ so that it now has a $50/50$ chance
of being in LHH or RHH.

\item  Atom $B$ observes atom $A$ and correlates itself to it, so that
either both occupy LHHs or both occupy RHHs. We make no assumptions about
how the observation is made.

\item  Depending on the state of $B$ we now compress $A$ to one of the two
halves; this involves no work, but the state of $B$ is now not known-it has a 
$50/50$ chance of being in LHH or RHH. Thus we have corrected $A$ at the
expense of randomizing $B$. It should be pointed out that by work, 
we always mean the work done by the atom (or on the atom) 
against the piston (or by the piston). 
As is usual in thermodynamical idealizations of this kind, all other works
are neglected (or assumed negligible). For example, the partition itself is assumed
to be very light (in fact, with zero weight), so that there is no work in pushing it.
Here there is no work done by the atom, since it is not contained in that part
of the box which is compressed (this information about the position of $A$ 
is recorded by $B$).

\item  In order to be able to repeat the error correction we need to reset $B$
to its initial state as in step 1. Thus we perform isothermal reversible compression
of $B$.
\end{enumerate}

\noindent Let us now analyze this process using entropy and free energy. In
step 1 both of the atoms possess $kT\log 2$ of free energy. After $A$
undergoes an error its free energy is decreased by $\Delta F_{A}=-kT\log 2,$
and nothing happens to $B.$ The total free energy is now $\Delta
F_{AB}=kT\log 2$. In step 3 the total free energy is still $\Delta
F_{AB}=kT\log 2,$ but the atoms are correlated. This means that atom $B$ has
information about $A$ (and vice verse). The amount of information is $k\log
2.$ This enables the error correction step to take place in step 4. This
does not change the total free energy, but the atoms are now decorrelated.
In step 5 a work of $kT\log 2$ is invested into resetting the state of atom $%
B$ so that the initial state in step 1 is reached. This completes the cycle
which can now start again. What happens to entropy? The entropy of each atom
is initially $0$. Then error increases the entropy of $A$ to $\Delta
S_{A}=k\log 2.$ In step 3 atoms get correlated so that they both have the
same entropy, i.e. $\Delta S_{B}=k\log 2.$ However, the crucial point is
that the total entropy does not change from step 2 to step 3. This is the
point of observation and the information gained by $B$ about the state of $A$
is $S_{A}+S_{B}-S_{AB}=k\log 2.$ In step 4 $\Delta S_{A}=-k\log 2$ and there
are no changes for atom $B.$ In the resetting step $\Delta S_{B}=-k\log 2,$
so that now both of the atoms have $0$ entropy like at the beginning.
Another change that took place, and this is the crux of Landauer's
principle, is that in the compression of atom $B$, work was invested and the
entropy of the environment increased by $k\log 2.$ This final entropy
increase is necessary for resetting and is in this case equal to the amount
of information gained in step 3. Landauer's principle of erasure states that
the entropy waste in resetting is at least as big as the information gain.
If this were not so, we could use the above cycle to do work by extracting
heat from the environment with no other changes and the Second Law of
thermodynamics (Kelvin's form) would be violated. Thus, here an error meant
that atom $A$'s ability to do work has been destroyed and in order to
correct this we needed another atom $B$ to transfer its free energy to $A.$
In this process atom $B$ loses its ability to do work and, in order to regain
it, an amount of $k\log 2$ of entropy has to be wasted (thus "saving" the
Second Law). To gain more familiarity with these kind of processes we will
now analyze quantum error correction in general settings and then apply our
reasoning to manipulations of entanglement.

\bigskip

\section{\noindent Quantum error correction as Maxwell's demon}

\bigskip

The aim of quantum error correction as presented in this section will be to
preserve a given quantum state of a quantum mechanical system (Knill $\&$ Laflamme 1997),
much as a refrigerator is meant to preserve the low temperature of food in a
higher temperature environment (room). Some work is performed on the
refrigerator which then reduces the entropy of food by increasing the
entropy of the surroundings. In accord with the Second law, the entropy
increase in the environment is at least as large as the entropy decrease of
the food. Analogously, when there is an error in the state of a quantum
system, then the entropy usually increases (this is, however, not always the
case as we will see later), and the error correction reduces it back to the
original state thereby decreasing the entropy of the system, but increasing
the entropy of the environment (or what we will call a garbage can). To
quantify this precisely let us look at error correction process in detail
(see also Nielsen \textit{et al.} 1998).

\bigskip

Suppose we wish to protect a pure state $\left| \psi \right\rangle
=\sum_{i}c_{i}\left| a_{i}\right\rangle $, where $\left\{ \left|
a_{i}\right\rangle \right\} $ form an orthonormal basis. This is usually
done by introducing redundancy, i.e. encoding the state of a system in a
larger Hilbert space according to some rule 
\[
\left| a_{i}\right\rangle \rightarrow \left| C_{i}\right\rangle 
\]
\noindent where $\left\{ \left| C_{i}\right\rangle \right\} $ are the so
called code-words. Note that this step was omitted in the classical case.
This is because the very existence of system $B$ can be interpreted as
encoding. The main difference between classical and quantum error correction
is that errors in classical
case can always be distinguished. In quantum mechanics these can lead to
non-orthogonal states so that the errors cannot always be distinguished and
corrected. So it might be said that the encoding in quantum mechanics
makes errors orthogonal and hence distinguishable (a precise mathematical statement of this is
given in Knill $\&$ Laflamme 1997). Of course, redundancy also exists in classical error
correction (above we have another system, $B$, to protect $A$), but the states are already orthogonal and distinguishable by the very nature of being classical. In the quantum case we will introduce an additional system,
called the apparatus, in order to detect different errors; this will play
the role that $B$ plays in classical error correction. Now the error
correction process can be viewed as a series of steps. First the initial
state is 
\[
\left| \psi _{c}\right\rangle \left| m\right\rangle \left| e\right\rangle 
\]

\noindent where $\left| \psi _{c}\right\rangle =\sum_{i}c_{i}\left|
C_{i}\right\rangle $ is the encoded state, $\left| m\right\rangle $ is the
initial state of the measuring apparatus and $\left| e\right\rangle $ is the
initial state of the environment. Now, the second stage is the occurrence of
errors, represented by the operators $\left\{ E_{i}\right\}$ which act on the state of the system only, after which we have
\[
\sum_{i}E_{i}\left| \psi _{c}\right\rangle \left| m\right\rangle \left|
e_{i}\right\rangle 
\]

\noindent Note that at this stage the measurement has not yet been made so
that the state of the apparatus is still disentangled from the rest. In general, the states
of the environment $\left| e_{i}\right\rangle$ need not be orthogonal (Vedral {\em et al.} 1997). If they are orthogonal this leads to a specific form of decoherence which we might call "dephasing" and will be analysed later. However, the formalism we present here is completely general and applies to any form of errors. Now the measurement occurs and we obtain 
\[
\sum_{i}E_{i}\left| \psi _{c}\right\rangle \left| m_{i}\right\rangle \left|
e_{i}\right\rangle 
\]

\noindent The error correction is seen as an application of $E_{i}^{-1}$,
conditional on the state $m_{i}$ (this, of course, cannot always be performed, but the code-words have to satisfy conditions in Knill $\&$ Laflamme 1997 in order to be correctable. Here we need not worry about this, our aim is only to understand global features of error correction). 
After this the state becomes
\[
\left| \psi _{c}\right\rangle \sum_{i}\left| m_{i}\right\rangle \left|
e_{i}\right\rangle 
\]

\noindent \noindent and the state of the system returns to the initial
encoded state; the error correction has worked. However, notice that the
state of the apparatus and the environment is not equal to their initial
state. This feature will be dealt with shortly. Before that let us note that
the total state (system+apparatus+environment) is always pure. Consequently the
von Neumann entropy is always zero. Therefore it is difficult to see how
this process will be compared to refrigeration where entropy is kept low at
the expense of the environment's entropy increasing. However, in general, the
environment is not accessible and we usually have no information about it
(if we had this information we would not need error correction!). Thus the
relevant entropies are those of the system and apparatus. This means that we
can {\em trace out} the state of the environment in the above picture; this
leads to dealing with mixed states and increasing and decreasing entropies.
In addition, the initial state of the system might be pure or mixed (above
we assumed a pure state) and these two cases we now analyze separately.

\subsection{\noindent Pure states}

\bigskip

We follow the above set of steps, but now the environment will be traced out
of the picture after errors have occurred. Therefore in the first step the
state is

\begin{enumerate}
\item  after errors $\sum_{i}E_{i}\left| \psi _{c}\right\rangle \left|
m\right\rangle \left| e_{i}\right\rangle ,$ where we assume the ''perfect''
decoherence, i.e. $\left\langle e_{j}|e_{i}\right\rangle =\delta _{ij}$, for simplicity.

\item  Now the environment is traced out leading to $\sum_{i}E_{i}\left|
\psi _{c}\right\rangle \left\langle \psi _{c}\right| E_{i}^{^{\dagger }}
\otimes 
\left| m\right\rangle \left\langle m\right| $. Note that this is not a
physical process, just a mathematical way of neglecting a part of the total
state (we introduce the direct product sign just to indicate separation
between the system and the apparatus; when there is no possibility of
confusion we will omit it).

\item  Then the system is observed, thus creating correlations between the
apparatus and the system $\sum_{i}E_{i}\left| \psi _{c}\right\rangle
\left\langle \psi _{c}\right| E_{i}^{^{\dagger }}\otimes 
\left| m_{i}\right\rangle \left\langle m_{i}\right| .$ We assume that the
observation is perfect so that $\left\langle m_{j}|m_{i}\right\rangle
=\delta _{ij}$; we will deal with the imperfect observation in the following
section. Note that we need different errors to lead to orthogonal states if
we wish to be able to correct them. Here, also, if the observation is imperfect then
error correction cannot be completely successful, since non-orthogonal states
cannot be distinguished with perfect efficiency (see also the discussion at the 
end of this section).

\item  The correction step happens and the system is decorrelated from the
apparatus so that we have $\left| \psi _{c}\right\rangle \left\langle \psi
_{c}\right| \otimes 
\sum_{i}\left| m_{i}\right\rangle \left\langle m_{i}\right| .$ As we
remarked before this is not equal to the initial state of the system and
apparatus. If we imagine that we have to perform correction a number of times in
succession, then this state of apparatus would not be helpful at all. We
need to somehow reset it back to the original state $\left| m\right\rangle .$

\item  This is done by introducing another system, called a garbage can
(gc), which is in the right state $\left| m\right\rangle ,$ so that the
total state is $\left| \psi _{c}\right\rangle \left\langle \psi _{c}\right| 
\otimes
\sum_{i}\left| m_{i}\right\rangle \left\langle m_{i}\right| 
\otimes 
\left| m\right\rangle \left\langle m\right| $, and then swapping the state
of the garbage can and the apparatus (this can be performed unitarily) so
that we finally obtain $\left| \psi _{c}\right\rangle \left\langle \psi
_{c}\right| \otimes 
\left| m\right\rangle \left\langle m\right| \otimes 
\sum_{i}\left| m_{i}\right\rangle \left\langle m_{i}\right| $. Only now the
system and the apparatus are ready to undergo another cycle of error
correction.

\noindent 
\end{enumerate}

\noindent We can now apply the entropy analysis to this error correction
cycle. In the first step the entropy of the system+apparatus has increased
by $\Delta S_{S+A}=S(\rho )$, where $\rho =$ $\sum_{i}E_{i}\left| \psi
_{c}\right\rangle \left\langle \psi _{c}\right| E_{i}^{^{\dagger }}$. Step 2
is not a physical operation and so there is no change in entropy. In the
third step, there is also no change in entropy; it is only that the
correlations between the system and the apparatus have been created. Step 4
is the same as the step 2, so no change in entropy on the whole. In step 5,
the entropy of the system+apparatus is zero since they are in the total
pure state. Thus, $\Delta S_{S+A}=-$ $S(\rho ),$ and now we see the formal
analogy with the refrigeration process: the net change in entropy of the
system+apparatus is zero, and the next error correction step can begin;
however, the gc has at the end increased in entropy by $\Delta S_{gc}=$ $%
S(\rho ).$ This is now exactly the manifestation of Landauer's erasure. The
information gain in step 3 is equal to the mutual entropy between the system
and the apparatus $I_{S+A}=S_{S}+S_{A}-S_{S+A}=S(\rho ).$ The logic behind
this formula is that before the observation the apparatus did not know
anything about the system, therefore system's state was uncertain by $S(\rho
),$ whereas after the observations it is zero - the apparatus knows
everything about the system. We note that this information is the Shannon mutual information which exists between the two (Schmidt) observables pertaining to the system 
and the apparatus. This needs to be erased at the end to start a
new cycle and the entropy increase is exactly (in this case) equal to the
information gained. So from the entropic point of view we have performed the
error correction in the most efficient way, since, in general, the gc entropy
increase is larger than the information gained (as we will see in subsection
C).

\bigskip

\noindent Next we consider correction of mixed states. This might at first
appear useless, because we might think that a mixed state is one that has
already undergone an error. This is, however, not necessarily so, and this
situation occurs when we are, for example, protecting a part of an entangled
bipartite system (see Vedral \textit{et al.} 1997). It might be thought that an analogous
case does not exist in classical error correction. There an error was
represented by a free expansion of the atom $A$ from step 1 to step 2.
However, we could have equally well started from $A$ occupying the whole
volume and treating an error as a "spontaneous" compression of the atom to one of the
halves. If $A$ was correlated to some other atom $C$ (so that they both occupied
LHH or RHH), then this compression would result in decorrelation which
really is an error. Thus classical and quantum error correction are in fact
very closely related which is also shown by their formal analogy to Maxwell's demon.

\subsection{\noindent Mixed states}

\bigskip

Now suppose that systems $A$ and $B$ are entangled and that we are only
performing error correction on the system $A$. Then this is in our case the same as
protecting a mixed state. We are not saying that protecting a mixed state is
in general the same as protecting entanglement. For example, if a state $|00\rangle+|11\rangle$
flips to $|01\rangle+|10\rangle$ with probability $1/2$, then the entanglement is destroyed, but  the reduced states of each subsystem are still preserved. What we mean is that quantum error correction is here developed to protect any pure state of a given system. In that case, any mixed state is also protected, and also any entanglement that it might have with some other systems. 
This also means that, using the standard quantum error correction (Knill $\&$ Laflamme 1997),  an entangled pair can be preserved just by protecting each of the subsystems separately.  
Now, for simplicity say that we have a mixture of two
orthogonal states $\left| \psi \right\rangle ,\left| \phi \right\rangle .$
The initial state is then without normalization (and without the system $B$)
\[
(\left| \psi \right\rangle \left\langle \psi \right| +\left| \phi
\right\rangle \left\langle \phi \right| )\otimes 
\left| e\right\rangle \left\langle e\right| \otimes 
\left| m\right\rangle \left\langle m\right| 
\]

\noindent And now we can go through all the above stages.

\begin{enumerate}
\item  \noindent error: $\sum_{i}E_{i}(\left| \psi \right\rangle
\left\langle \psi \right| +\left| \phi \right\rangle \left\langle \phi
\right| )E_{i}^{^{\dagger }}\otimes 
\left| e_{i}\right\rangle \left\langle e_{i}\right| \otimes 
\left| m\right\rangle \left\langle m\right| ;$

\item  tracing out the environment: $\sum_{i}E_{i}(\left| \psi \right\rangle
\left\langle \psi \right| +\left| \phi \right\rangle \left\langle \phi
\right| )E_{i}^{^{\dagger }}\otimes 
\left| m\right\rangle \left\langle m\right| ;$

\item  observation: $\sum_{i}E_{i}(\left| \psi \right\rangle \left\langle
\psi \right| +\left| \phi \right\rangle \left\langle \phi \right|
)E_{i}^{^{\dagger }}\otimes 
\left| m_{i}\right\rangle \left\langle m_{i}\right| ;$

\item  correction: $(\left| \psi \right\rangle \left\langle \psi \right|
+\left| \phi \right\rangle \left\langle \phi \right| )\otimes 
\sum_{i}\left| m_{i}\right\rangle \left\langle m_{i}\right| ;$

\item  resetting: $(\left| \psi \right\rangle \left\langle \psi \right|
+\left| \phi \right\rangle \left\langle \phi \right| )\otimes 
\sum_{i}\left| m_{i}\right\rangle \left\langle m_{i}\right|\otimes 
\left| m\right\rangle \left\langle m\right| \\ \rightarrow (\left| \psi
\right\rangle \left\langle \psi \right| +\left| \phi \right\rangle
\left\langle \phi \right| )\otimes 
\left| m\right\rangle \left\langle m\right| \otimes
\sum_{i}\left| m_{i}\right\rangle \left\langle m_{i}\right| .$
\end{enumerate}

\bigskip

\noindent The entropy analysis is now as follows. In step 1, $\Delta
S_{S+A}=S(\rho _{f})-S(\rho _{i})$, where $\rho _{f}=\sum_{i}E_{i}(\left|
\psi \right\rangle \left\langle \psi \right| +\left| \phi \right\rangle
\left\langle \phi \right| )E_{i}^{^{\dagger }}$ and $\rho _{i}=\left| \psi
\right\rangle \left\langle \psi \right| +\left| \phi \right\rangle
\left\langle \phi \right| $ (not normalized)$.$ In steps 2 and 3 there is no
change of entropy, although in step 3 an amount of $I=S(\rho _{f})$
information was gained if the correlations between the system and the
apparatus are perfect (i.e. $\left\langle m_{j}|m_{i}\right\rangle =\delta
_{ij})$. In step 4, $\Delta S_{S+A}=S(\rho_{i})$ as the system and the apparatus become
decorrelated. In step 5, $\Delta S_{S+A}=-S(\rho _{f}),$ but the entropy of the gc
increases by $S(\rho _{f}).$ Thus altogether $\Delta S_{S+A}=0$, and the
entropy of the gc has increased by exactly the same amount as the
information gained in step 3 thus confirming Landauer's principle again.

\noindent Now we want to analyze what happens if the observation in step 3
is imperfect. Suppose for simplicity that we only have two errors $E_{1}$
and $E_{2}.$ Then there would be only two states of the apparatus $\left|
m_{1}\right\rangle $ and $\left| m_{2}\right\rangle $; an imperfect
observation would imply that $\left\langle m_{1}|m_{2}\right\rangle =a>0.$
Now the entropy of information erasure is $S(\left| m_{1}\right\rangle
\left\langle m_{1}\right| +\left| m_{2}\right\rangle \left\langle
m_{2}\right| )$ which is smaller than when $\left| m_{1}\right\rangle $ and $%
\left| m_{2}\right\rangle $ are orthogonal. This implies via Landauer's
principle that the information gained in step 3 would be smaller than when
the apparatus states are orthogonal and this in turn leads to imperfect
error correction. Thus, doing perfect error correction without perfect
information gain is forbidden by the Second law of thermodynamics via
Landauer's principle. This is analogous to von Neumann's (1952) proof that being
able to distinguish perfectly between two non-orthogonal states would
lead directly to violation of the Second Law of thermodynamics.
\ 

\bigskip

\subsection{General erasure}

\bigskip

Previously we described erasure as a swap operation between the gc and the
system. Now we will describe a more general way of erasing information, but
which will be central to our understanding of entanglement manipulations in
section 4. We follow Lubkin's (1987) analysis in somewhat more general settings.

\bigskip

A more general way of conducting erasure (resetting) of the apparatus is to
assume that there is a reservoir which is in thermal equilibrium in a Gibbs
state at certain temperature $T$. To erase the state of the apparatus we just
throw it into the reservoir and bring in another pure state. The entropy
increase of the operation now consists of two parts: the apparatus reaches
the state of the reservoir and this entropy is now added to the reservoir
entropy, and also the rest of the reservoir changes its entropy due to this
interaction which is the difference in the apparatus internal energy before
and after the resetting (no work is done in this process). This quantum
approach to equilibrium was also studied by Partovi (1989). A good model is obtained by 
imagining that 
the reservoir consists of a great number of systems (of the same "size" as the 
apparatus) all in the same quantum 
equilibrium state $\omega$. Then the apparatus, which is in some state $\rho$, 
interacts with these reservoir systems one at a time. Each time there is an 
interaction, the state of the apparatus approaches more closely the 
state of the reservoir, while that single reservoir
system also changes its state away from the equilibrium. However, the systems
in the bath are numerous so that after a certain number of collisions the
apparatus state will approach the state of the reservoir, while the reservoir
will not change much since it is very large (this is equivalent to the so called Born-Markov 
approximation that leads to irreversible dynamics of the apparatus described here). 

Bearing all this in mind, we now reset the apparatus by plunging it into a
reservoir in a thermal equilibrium at temperature $T.$ Let the state of the
reservoir be
\[
\omega =\frac{e^{-\beta H}}{Z}=\sum_{j}q_{j}\left| \varepsilon
_{j}\right\rangle \left\langle \varepsilon _{j}\right| 
\]

\noindent where $H=$ $\sum_{i}\varepsilon _{i}\left| \varepsilon
_{i}\right\rangle \left\langle \varepsilon _{i}\right| $ is the Hamiltonian of the reservoir,
$Z=tr(e^{-\beta H})$ is the partition function and $\beta^{-1}=k T$, where 
$k$ is the Boltzmann constant. Now suppose that due to
the measurement the entropy of the apparatus is $S(\rho )$ (and an amount $
S(\rho )$ of information has been gained)$,$ where $\rho
=\sum_{i}r_{i}\left| r_{i}\right\rangle \left\langle r_{i}\right| $ is the
eigen expansion of the apparatus state. Now the total entropy increase in
the erasure is (there are two parts as we argued above: 1. change in the entropy
of the apparatus and 2. change in the entropy of the reservoir)
\[
\Delta S_{er}=\Delta S_{app}+\Delta S_{res} 
\]

\noindent We immediately know that $\Delta S_{app}=S(\omega )$, since the
state of apparatus (no matter what state it was before) is now erased to be the same as
that of the reservoir. On the other hand, the entropy change in the
reservoir is the average over all states $\left| r_{i}\right\rangle $ of
heat received by the reservoir divided by the temperature. This is minus the
heat received by the apparatus divided by the temperature; the heat received
by the apparatus is the internal energy after the resetting minus initial internal
energy $\left\langle r_{i}\right| H$ $\left| r_{i}\right\rangle .$ Thus,
\begin{eqnarray*}
\Delta S_{res} &=&-\sum_{k}r_{k}\frac{tr(\omega H)-\left\langle r_{k}\right|
H\left| r_{k}\right\rangle }{T} \\
&=&\sum_{k}(r_{k}\sum_{j}|\left\langle r_{k}|\varepsilon _{j}\right\rangle
|^{2}-q_{k})(-\log q_{k}-\log Z) \\
&=&-tr(\rho -\omega )(\log \omega -\log Z)\\
&=& tr(\omega-\rho)\log \omega
\end{eqnarray*}

\noindent Altogether we have that
\[
\Delta S_{er}=-tr(\rho \log \omega ) 
\]

\noindent (This result generalizes Lubkin's (1987) result which applies 
only when $\left[ \rho ,\omega \right] =0$). In general, however, the
information gain is equal to $S(\rho )$, the entropy increase in the
apparatus. Thus, we see that
\[
\Delta S_{er}=-tr(\rho \log \omega )\geq S(\rho )=I 
\]

\noindent and Landauer's principle is confirmed (the inequality follows from
the fact that the quantum relative entropy $S(\rho ||\omega )=-tr(\rho \log
\omega )-S(\rho )$ is non-negative). So the erasure is least wasteful when $%
\omega =$ $\rho ,$ in which case the entropy of erasure is equal to $S(\rho
),$ the information gain. This is when the reservoir is in the same state as the
state of the apparatus we are trying to erase. In this case we just have a 
state swap between the new pure state of the apparatus which is used to replace our old
state $\rho$. This, in fact, was the case in all our examples of error
correction above. However, sometimes it is impossible to meet this condition
and it is this case we turn to next.

\bigskip

\section{Entanglement purification}

\bigskip

\subsection{General Considerations}

Entanglement purification is a procedure whereby an ensemble of bipartite
quantum systems all in a state $\rho $ is converted to an subensemble of
pure maximally entangled states by local operation on the systems separately
and with the aid of classical communications (Bennett \textit{et al.} 1996$a$). 
The rest of the
pairs end up completely separable (i.e. disentangled) and can be 
taken to be in a pure state of
the form $\left| \psi \right\rangle \left| \phi \right\rangle $. For the
sake of simplicity let us first assume that the initial ensemble is in a
pure, but not maximally entangled state. To link this with our previous
analysis let us see how this situation might arise. Let a system $S$ be in
the state
\[
\left| \psi _{S}\right\rangle =\frac{1}{\sqrt{N}}\sum_{i=1}^{N}\left|
s_{i}\right\rangle 
\]

\noindent where $\left\{ \left| s_{i}\right\rangle \right\} $ is an
orthonormal basis. Let now the apparatus observe this superposition after
which the state is
\[
\left| \psi _{S+A}\right\rangle =\frac{1}{\sqrt{N}}\sum_{i=1}^{N}\left|
s_{i}\right\rangle \left| a_{i}\right\rangle 
\]

\noindent but let the observation be imperfect so that $\left\{ \left|
a_{i}\right\rangle \right\} $ is NOT an orthogonal set (which means that $S$
and $A$ are not maximally entangled). However, suppose that by acting on the
apparatus we can transform the whole state $\left| \psi _{S+A}\right\rangle$
into the maximally entangled state
\[
\left| \phi _{S+A}\right\rangle =\frac{1}{\sqrt{N}}\sum_{i=1}^{N}\left|
s_{i}\right\rangle \left| b_{i}\right\rangle 
\]

\noindent where $\left\{ \left| b_{i}\right\rangle \right\} $ IS an
orthogonal set. This does not increase the information between the apparatus
and the system since we are not interacting with the system at all. The
crucial question is: what is the probability with which we can do this? Let
\[
tr_{S}(\left| \psi _{S+A}\right\rangle \left\langle \psi _{S+A}\right|
)=\rho _{A} 
\]

\noindent be the state of the apparatus after the perfect measurement. Then
Landauer's principle says that the entropy of erasure is $S(\rho _{A})$ and
this has to be greater than or equal to the information gain. But let us
look at what the information gain is after we purified the state to $\left|
\phi _{S+A}\right\rangle $ with a probability $p.$ First of all, we gained $%
p\log N$ information about the system since we have maximal correlations
now. Secondly the rest of the state contains no information (we assume it is
completely disentangled) and this is with probability $(1-p).$ Thus writing
Landauer's principle leads to
\[
S(\rho _{A})\geq p\log N 
\]

\noindent The upper bound to purification efficiency is therefore
\[
p\leq S(\rho _{A})/\log N 
\]

\noindent That the upper bound is achievable was shown by Bennett et al (1996$a$). 
We stress that in this reasoning we used that the entropy erasure
is greater than or equal to the information gain before purification (by
Landauer), and this in turn is greater than or equal to the information
after purification (since, in purification, the apparatus does not interact
with the system).

\bigskip

Let us now analyze when the initial state of the system and the apparatus is
mixed in a state $\rho $. Then by Lubkin's method the entropy of erasure is
\[
\Delta S_{er}=-tr(\rho \log \omega ) 
\]

\noindent The information gain after the purification is
\[
I=S(\rho )+p\log N 
\]

\noindent where $S(\rho )$ comes from the fact that all the states after
purification are pure (being either maximally entangled or disentangled), so that
the total information gain is the uncertainty before (i.e. $S(\rho )$) minus the
uncertainty after (i.e. zero). We can view this in yet another way. The fact that 
$\rho$ is mixed means that it is entangled to another system, which we call ancilla. 
The total state of system+apparatus+ancilla can always be chosen to be pure. Now, system+apparatus are correlated to the ancilla, 
since the system+apparatus itself is in a mixed state 
(the ancilla is also mixed and has the same eigen-spectrum as system+apparatus
which means that it also has the same von Neumann entropy). After the purification this correlation disappears, and the resulting information is $S(\rho ).$
Using Landauer's principle now gives
\[
-tr(\rho \log \omega )\geq S(\rho )+p\log N 
\]

\noindent or
\[
p\leq S(\rho ||\omega )/\log N 
\]

\noindent where $S(\rho ||\omega )=-tr(\rho \log \omega )-S(\rho )$ is
quantum relative entropy which we met in the previous section. Now a tight
upper bound on purification is found when $S(\rho ||\omega )$ is minimal.
However, remembering that we now acted on the system and the apparatus
separately (i.e. all the operations were local), the state of the reservoir
for resetting will also have to be local (i.e. separable or disentangled 
as in Vedral $\&$ Plenio 1998). This means that the system is reset by plunging it into
its own reservoir and the apparatus is reset by plunging it into its own, but separate, reservoir.
By reservoir we will always mean two of these reservoirs together unless stated otherwise. 
We can always assume that they are in a separable, but classically correlated state 
as will be shown later in this section. Let us call the set of all separable states 
of the system and the apparatus ${\it D}.$ Then,
\[
p\leq \text{min}_{\omega \in {\it D}}S(\rho ||\omega )/\log N 
\]

\noindent The quantity $E_{RE}(\rho )=$min$_{\omega \in {\it D}}S(\rho
||\omega )$ is known as the relative entropy of entanglement and has been
recently argued by Vedral and Plenio (1998) and separately Rains (1998) 
to be an upper bound on the efficiency of purification procedures.
For $\rho $ pure this reduces to our previous result. These results were
originally derived from the principle that entanglement, being a non-local
property, cannot increase under local operations and classical
communications. This implied finding a mathematical form of local operations
and then finding a measure which is non-increasing under them (Vedral $\&$ Plenio 1998).
Now, we have derived the same result in a simpler and more physical way, but
from at first sight completely different direction, by taking Landauer's
principle. We have thus related the ''no local increase of entanglement''
principle to the Second Law of thermodynamics. Let us briefly summarize this
link. Local increase of entanglement is in this context equivalent to {\it
perfectly} distinguishing between non-orthogonal states. If we have this
ability we can then violate the Second law with Maxwell's demon that Bennett (1982)
used in his analysis (also von Neumann (1952) showed this by a
different argument). Therefore, local increase of entanglement
is seen to be prohibited by the Second Law. The additional principle that
we used is that information about some system can only be gained if we
interact with it. We emphasize that this principle is not necessarily
related only to entanglement. We saw in the previous section that, in error
correction, the apparatus correlates itself with the system in order to
correct errors and these correlations are purely classical in nature. It is
here, however, that the link between ''no local increase of entanglement''
and Landauer's principle is most clearly seen: entanglement between two
subsystems cannot be increased unless one subsystem gains more information
about the other one, but this cannot be done locally without interaction.

It should be noted that the reservoirs in general have to be classically correlated 
(in order to obtain a tight upper bound on purification; otherwise, the bound still holds, but
is in general too high).
This poses a question as to how this could be achieved in practice. A way to do that
is to remember that this result is applicable in the asymptotic case, meaning that
we can average over a large number of different, but uncorrelated reservoir states,
which are certainly natural to consider. So, if $\omega=\sum_{i} p_i \omega_{S}^{i}\otimes 
\omega_{A}^{i}$ is the state of the reservoir achieving the minimum (where $S$ refers to the reservoir into which the system is immersed and $A$ to the reservoir into which the apparatus is immersed to erase their mutual information), then this implies that we would
be using reservoirs of the form $\omega_{S}^{i}\otimes \omega_{R}^{i}$ with frequencies
$p_i$ and this would on average produce the state $\omega$. Therefore if we consider
at $n$ initial mixed states of system+apparatus, then to delete these correlations, we use $p_{1}n$ times the uncorrelated
reservoir state $\omega _{S}^{1}\otimes \omega _{A}^{1},$ $p_{2}n$ times the
uncorrelated state $\omega _{S}^{2}\otimes \omega _{A}^{2}$ and so on. Note that while each
of these reservoirs is in a thermal state with a well defined temperature, the total
(classically correlated) state does not have a well defined temperature. In general, however,
the bound is universal and holds no matter how this purification is performed as long as it is local in character. Also note that if we are allowed to have a common reservoir for the 
system and the apparatus, then the erasure can be different to the above, i.e. it can be 
smaller than the above local erasure. However, if we allow nonlocal operations, then we can
also create additional entanglement and the above analysis would not apply. An alternative way to view this local erasure is to allow a common reservoir for the system and the apparatus, but
to restrict its thermal states to separable states only. Then we are sure that no additional
entanglement is created to the already existing between the system and the apparatus and
all our results remain true.    

\bigskip

This now leads us to state another, equivalent way of interpreting the relative
entropy of entanglement. Suppose that we again have a disentangled bipartite
system+apparatus, but this time in a thermal state $\omega =e^{-\beta H}/Z$, where $\beta^{-1}=kT$, $k$
is the Boltzmann constant, $T$ the temperature, $H$ is the Hamiltonian and $Z$ the 
partition function (note that this is now the state of system+apparatus which is the 
same as the state of the reservoir used for resetting. Here, however, there are
no further locality requirements and we do have 
a well defined temperature). The question that we wish to ask now is how much free energy would this system gain by going to the entangled state $\rho $ (see Donald 1987
for applications of quantum relative entropy to statistical mechanics, and 
Partovi 1989 for the quantum basis of thermodynamics)? Note that here we allow
any operations, and are not restricted by locality requirements. First of all, the free
energy is given by
\[
F=U-TS 
\]

\noindent so that 
\begin{eqnarray*}
F(\rho ) &=&tr(\rho H)-\beta ^{-1}S(\rho ) \\
F(\omega ) &=&-\beta ^{-1}\log Z
\end{eqnarray*}

\noindent Therefore the difference in free energies of $\rho $ and $\omega $
is
\[
F(\rho )-F(\omega )=\beta ^{-1}S(\rho ||\omega ) 
\]

\noindent In the light of this, {\it entanglement as given by the relative
entropy would be proportional to the amount of free energy lost in deleting
all the quantum correlations and creating solely classical correlations} by
e.g. plunging the system+apparatus into a reservoir whose thermal state is given by
that classically correlated state (now we do allow a common reservoir with a
well defined thermal state and temperature). The constant of proportionality is the
temperature times the Boltzmann constant $k$.

\bigskip

\noindent We stress that entanglement of creation, another measure of
quantum correlations introduced by Bennett et al (1996$b$), can also be
interpreted using above methods. It is also an upper bound to the efficiency
of purification procedures although it is not tight as it is larger than the
relative entropy of entanglement (Vedral $\&$ Plenio 1998). It arises from applying
Landauer's principle to each pure state in the decomposition of $\rho
=\sum_{i}p_{i}\left| \psi _{i}\right\rangle \left\langle \psi _{i}\right| $
(note that this is not necessarily the eigen-decomposition)$.$ In fact, we can define
the entanglement of creation via the free energy as {\em the average
decrease in free energy due to resetting each of }$\left| \psi
_{i}\right\rangle =\sum_{j}c_{ij}\left| a_{j}\right\rangle \left|
b_{j}\right\rangle ${\em \ to a completely separable state }$\omega
_{i}=\sum_{i}|c_{ij}|^{2}\left| a_{j}\right\rangle \left| b_{j}\right\rangle
\left\langle a_{j}\right| \left\langle b_{j}\right| $ (strictly speaking,
the entanglement of creation is actually the minimum of this quantity over
all possible decompositions of $\rho $). Mathematically this can be
expressed as
\[
E_{C}=\frac{\min_{\text{decomp of }\rho }\sum_{i}\Delta F_{i}}{T} 
\]
where $\Delta F_{i}=\beta ^{-1}S(\left| \psi _{i}\right\rangle ||\omega
_{i}) $ (which is equal to the von$\;$Neumann reduced entropy of $\left|
\psi _{i}\right\rangle $ Vedral $\&$ Plenio 1998).

At the end of this section we emphasize again why local operations were
important when we wanted to derive a bound on purification of entanglement. 
First of all, it is believed that entanglement does not increase
under local operations and classical communication. Here, however, we did
not use this fact to derive bounds on the efficiency of purification
procedures. We only used the fact that local operations on one subsystem
(e.g. apparatus) do not tell us anything more about some other subsystem
than is already known; in other words, if the operations were not local the
above analysis would be wrong. In fact, the upper bounds we derived on
entanglement purification can now be restated: the most successful
purification is the one which wastes no information, i.e. the free energy
needed to reset the state of the ensemble before purification should be
equal to the free energy needed to reset the state of the ensemble after
purification (i.e. Landauer's erasure bound is saturated). At the end of this 
section we showed an equivalent way of interpreting the relative entropy of
entanglement where the operations are not restricted to the local ones only, 
but the reservoir state still has to be disentangled.  

\subsection{\noindent Ensemble versus single trial view}

\bigskip

All the results we considered above are actually appropriate from the
ensemble point of view. To explain what we mean by this consider the
following problem. Suppose that Alice and Bob share a {\em single }pair of
entangled systems in a state
\[
\left| \psi \right\rangle =a\left| 0\right\rangle \left| 0\right\rangle
+b\left| 1\right\rangle \left| 1\right\rangle 
\]

\noindent Then it can be proven (Lo $\&$ Popescu 1997) that the best efficiency (i.e.
the highest probability) with which they can purify this state by acting
locally on their own systems is $2b^{2}$ if $b<a$. However, $%
2b^{2}<-a^{2}\log a^{2}-b^{2}\log b^{2}$ (the equality is achieved only when 
$b^{2}=1/2,$ and the state is already maximally entangled) which is the
efficiency we derived above from Landauer's erasure principle. Thus
Landauer's efficiency limit is only reached asymptotically when Alice and
Bob share and infinite ensemble of entangled systems and they operate on all
of their particles at the same time. So, although Landauer's erasure holds
true even if we operate on single pairs (a ''single shot'' measurement), it
gives an overestimate of the efficiency of this process. We note that, strictly speaking, 
both of these problems involve ensembles, but in the single shot view we are allowed to
act on only one pair at a time, whereas otherwise we can act on all of them simultaneously. It is therefore no surprise that the former method, which is a special case of the latter method, is less efficient. This is the reason why the bound we presented in this paper is too high for the single shot purification. On the whole, however, by performing erasure the way we imagined, where we use thermal reservoirs to delete information, we have derived a universal upper bound no matter how
the purification is performed. Thus an open
question would be whether it is more appropriate to use a different measure
for erasure in the single shot case since the amount of entanglement as measured by the
purification efficiency is different for a single pair measurement and for
an ensemble measurement. This would be important to consider, since most
of the practical manipulations at present involve only a few entangled
particles and, as we said, the entropic measures overestimate various
efficiencies of entanglement processing. So, in our bounds on entanglement
purification, $\rho $ should actually stand for $\rho ^{n}=\rho \otimes 
\rho ...\otimes \rho $ ($n$ times), where each $\rho $ now refers to a
single pair of quantum systems (also $\log N$ would become $n\log N$). This
brings us to the question of whether min$_{\omega \in {\it D}}S(\rho
^{n}||\omega )=n$min$_{\omega \in {\it D}}S(\rho ||\omega ),$ i.e. whether
the relative entropy of entanglement is additive, which is still open
(although see Rains 1998). A reasonable conjecture is that in all the
quantum information manipulations that involve large ensembles the above
reasoning will be suitable. Examples of this are quantum data compression 
(Schumacher 1995) and capacity of a quantum channel (see, for example, Feynman's (1996)
derivation of Shannon's coding for classical binary symmetric
channels using Landauer's principle). In quantum data compression, for
instance, the free energy lost in deleting before compression is $n\beta
^{-1}S(\rho )$ and after the compression is $m\beta ^{-1}\log N.$ These two
free energies should be equal if no information is lost (i.e. if we wish to
have maximum efficiency) in compressing and therefore $m/n=$ $S(\rho )/\log
N$ as shown by Schumacher 1995. 

\bigskip

\section{\noindent Conclusions}

\bigskip

We have analyzed general classical and quantum error correcting procedure
from the entropic perspective. We have shown that the amount of information
gained in the observation step needed to perform correction is then turned
into an equal amount of wasted entropy in a gc. This gc is needed to reset
the apparatus to its initial state so that the next cycle of error
correction can then be performed. This fact is equally true both in the
classical the quantum case and is known as Landauer's principle of
information erasure. We then analyzed purification procedures using the same
principle. Surprisingly, Landauer's principle when applied appropriately
yields the correct upper bounds to the efficiency of purification
procedures. Whether these bounds can be achieved in general remains an open
question. Landauer's principle therefore provides a physical basis for
several entanglement measures, notably relative entropy of entanglement and
the entanglement of creation. In addition this provides a link between the
principle of ''no local increase of entanglement'' and the Second Law of
thermodynamics. Further open questions are the implications of Landauer's
erasure to other forms of quantum information manipulations such as quantum
cloning.

\bigskip

\noindent {\bf Acknowledgments. }I would like to thank Guido Bacciagaluppi
for inviting me to give a talk at the Philosophy department in Oxford which
stimulated me to think more about the relationship between thermodynamics
and entanglement thus resulting in this exposition. I would also like to thank
M. Plenio and B. Schumacher for useful comments and discussions on the subject of this 
paper.

\bigskip

\bigskip

\noindent Fig 1. Classical error correction as a Maxwell's demon. Steps are
detailed in the text and their significance explained: (1) states of atoms A
and B are initially uncorrelated; (2) atom A undergoes an error; (3) atom B
observes the atom A, and the atoms thereby become correlated; (4) atom A is
corrected to its initial state and the atoms are now uncorrelated; (5) atom
B is returned to its initial state and the whole cycle can start again.

\end{document}